\documentclass[11pt,a4paper]{article}
\usepackage{jheppub_kim}

\usepackage{pdflscape}
\usepackage{amsmath}
\usepackage{amssymb}
\usepackage{dcolumn}
\usepackage{bm}
\usepackage{color}
\usepackage{epsfig}
\usepackage{amsfonts}
\usepackage{graphicx}
\usepackage{subfigure}

\newcommand{\be}{\begin{equation}}
\newcommand{\ee}{\end{equation}}
\newcommand{\bea}{\begin{eqnarray}}
\newcommand{\eea}{\end{eqnarray}}

\def\be{\begin{equation}}
\def\ee{\end{equation}}
\def\bea{\begin{eqnarray}}
\def\eea{\end{eqnarray}}

\begin{document}

\title{Holographic real and  imaginary potentials of heavy quarkonium in Yang-Mills-dilaton black holes}
\author[a]{M. Kioumarsipour,}
\author[b]{B. Khanpour}

\affiliation[a]{Sciences Faculty, Department of Physics, University of Mazandaran, Babolsar, Iran}

\affiliation[b]{Department of Electrical Engineering, Mazandaran University of Science and Technology, Babol, Iran}

\emailAdd{ma.kioumarsi@gmail.com}
\emailAdd{khanpourbh@gmail.com}

\abstract{\\
Within the methods of the real and the imaginary potential for investigating the melting of a heavy quarkonium in the AdS/CFT correspondence, we study the effects of the dynamical exponent, the hyperscaling violation parameter, and the Yang-Mills charge on both methods in the hyperscaling violated Yang-Mills-dilaton metric background. In the real part of the potential, we find out that the dynamical exponent, the hyperscaling violation parameter, and the Yang-Mills charge increase the real potential and decrease the dissociation length, and hence the quarkonium dissociates easily. In the imaginary part of the potential, we show that the dynamical exponent, the hyperscaling violation parameter, and the Yang-Mills charge tend to decrease the thermal width, thus strengthening the suppression of the heavy quarkonium, and making the melting process easier. 
Also, we show that the obtained results from the real and imaginary potential methods are compatible and complementary. \\
\textbf{Keywords:} AdS/CFT correspondence, Real Potential, Imaginary potential,  Hyperscaling violation.}

\maketitle

\section{Introduction}
The AdS/CFT correspondence or the gauge/gravity duality is a conjectured idea for better understanding the unknown parts of nature that relates theories of gravity in $ d+1 $ dimensions Anti-de Sitter (AdS) spacetimes to $ d $ dimensional strongly coupled gauge field theories with conformal symmetry (CFT) where live on the boundary of the AdS spacetimes. This powerful tool attracted a lot of attention. One of the most important applications of this duality is the study and analysis of different quantities in Quantum chromodynamics (QCD). In Relativistic Heavy Ion Collider (RHIC) and the Large Hadron Collider (LHC), where heavy ions were collided with each other, a new phase of matter called quark-gluon plasma (QGP) appeared \cite{q,g,p} which was strongly coupled, and impossible to be described with perturbation techniques. Since the gauge/gravity duality is a powerful and practical non-perturbative technique, it can be applied to investigate the properties of QGP. \cite{a,d,s}. 
  
The original prescription of the gauge/gravity duality relates the type IIB string theory on $ AdS_5 \times S^5 $ spacetime to the four-dimensional $\mathcal{N}=4 $ supersymmetry Yang-Mills (SYM) field theory \cite{a}. Since at zero temperature, $\mathcal{N}=4 $ SYM theory has supersymmetry and conformal invariance, it can not exactly correspond to QCD. While these theories indicate similar non-Abelian strongly coupled plasmas at high temperatures, at least above the critical temperature. The field theory which is dual to the AdS geometry is a conformal invariance theory. Indeed, the real QCD is not conformally invariant, therefore the conformal invariance of AdS spacetime should be broken somehow. For this reason to reach a more realistic description of QCD, the AdS/CFT correspondence should be modified. Some techniques that can be used to modify the AdS/CFT correspondence are the hard wall \cite{h1} and the soft wall \cite{s1} models. 
Another remarkable modification of the AdS/CFT correspondence is its extension to the anisotropic Lifshitz spacetime \cite{Azeyanagi:2009pr},
\begin{equation}
	ds^{2}= r^{2z}\left(-dt^{2}+\sum_{i=1}^{p} dx_{i}^{2}\right) +r^2\sum_{j=p+1}^{d}dy_{j}^{2}+\dfrac{dr^{2}}{r^{2}},~~~~~~~~0\leq p\leq d-1. \label{tt1}
\end{equation}
The dynamical exponent $z(\neq 1)$ indicates the degree of the deviation of Lorentz symmetry and anisotropy. Under the scaling $(t,x_{i},y_{j},r)\rightarrow(\lambda^z t,\lambda^z x_{i},\lambda y_{j},\frac{r}{\lambda}), $ this metric is invariant. In the particular case $p=0$, this metric exhibits non-relativistic spacetime where is scaled anisotropically in the time direction. The metric (\ref{tt1}) can be rewritten as the following form,
\begin{equation}
	ds^{2}= \rho^{2}\left(-dt^{2}+\sum_{i=1}^{p} dx_{i}^{2}\right) +\rho^{\frac{2}{z}}\sum_{j=p+1}^{d}dy_{j}^{2}+\dfrac{d\rho^{2}}{\rho^{2}}, \label{tt2}
\end{equation}    
by a coordinate transformation $\rho=r^{z}$.
The above metric displays invariance under an anisotropic scaling transformation $(t,x_{i},y_{j},\rho)\rightarrow(\lambda t,\lambda x_{i},\lambda^{\frac{1}{z}} y_{j},\frac{\rho}{\lambda}).$ Here, the directions of $y_{j} $ cause the Lorentz symmetry violation and anisotropy. In the cases $1\leq p\leq d-1$, the metric (\ref{tt2}) represents a space-like anisotropic metric \cite{Azeyanagi:2009pr}. Some anisotropic backgrounds were investigated in refs \cite{Mateos:2011ix,Mateos:2011tv,Giataganas:2012zy,Giataganas:2013lga}. As argued in \cite{Azeyanagi:2009pr}, a space-like anisotropic background may be equivalent to (\ref{tt2}) with its own special $z$, like as z=3/2 in that paper.   

A technique for breaking the conformal invariance of the AdS spacetime is adding the scalar and gauge fields to the Einstein gravity  \cite{l1,l3,l4,l5,AA,AB}. This leads to the appearance of two parameters: the dynamical exponent $ z $ and the hyperscaling violation parameter $ \theta $,
\begin{equation}
	ds^{2}=r^{-2\theta /d}\left( -r^{2z}dt^{2}+\dfrac{dr^{2}}{r^{2}}+r^{2}d{\bf x}^{2}\right). \label{tt}
\end{equation}
 
The dynamical exponent $ z $ scales the space and time directions differently and 
the hyperscaling violation parameter $ \theta $ breaks the conformal invariance. However, such metrics exhibit the scale transformations as follows,    
\begin{equation}
	t\rightarrow\lambda^{z}t,~~~~~r\rightarrow\lambda^{-1}r,~~~~~x_{i}\rightarrow\lambda x_{i},~~~~~ds_{d+2}\rightarrow\lambda^{\theta/d}ds_{d+2}.
\end{equation}
There are different ways to make a metric containing both the dynamical exponent $ z $ and the hyperscaling violation parameter $ \theta $, such as: Einstein-Maxwell-dilaton (EMD) system \cite{alii}, Einstein-Maxwell-dilaton gravity with $ R^{2} $ corrections \cite{1312.2261}, Einstein-Proca-dilaton model \cite{1212.3263}, Einstein-Maxwell-dilaton-axion model \cite{1606.07905}, cubic gravity \cite{1508.05614}, Ho\v{r}ava-Lifshitz gravity \cite{1212.4190} and non-supersymmetric truncations of $ \omega $-deformed $ \mathcal{N}=8 $ gauged supergravity \cite{1411.0010}.  
The references \cite{1209, 1205.0412, 1404.5399, 1608.03247, 1608.04394, 1807.09770, tay} are also useful in this regard. The Lifshitz scaling and hyperscaling violation geometries have typically a running dilatonic scalar \cite{arXiv:1005.4690,arXiv:1107.2116,arXiv:0905.3337}. The running scalar causes the solutions not to be trusted in the deep IR, i.e. they are IR incomplete \cite{arXiv:1305.3279,arXiv:1208.1752}.
Genuine null IR singularities with diverging tidal forces may be generated by choosing generic $ z $ and $ \theta $  \cite{AA,o1,o2}. These singularities are removed by applying stringy effects \cite{o3,o5}.  Because these solutions do not present a valid expression of the geometry in the deep IR, one has to modify them. In other words, the RG flow connects two different spacetimes, the AdS spacetimes in the UV to a non-relativistic spacetime with the dynamical and the hyperscaling violation exponents in the IR. RG flows from UV to IR with generic $ z $ and $ \theta $, and novel RG flows, the so-called \textit{boomerang RG flows} are described in Refs \cite{1708,o8}. In fact, the Lifshitz and hyperscaling violation geometries can be expressed as an effective holographic explanation of some strongly coupled QFT which are given below a specific energy scale of the theory. In other words, metrics such as (\ref{tt}) do not show a good expression in all distances of $ r $. Therefore, they should be considered on a finite $ r=r_{F} $ cutoff, which means that the dual field theory is placed on this slice \cite{null}. Also, hyperscaling violation theories play an important role in studying the condensed matter, the holographic description of Fermi surfaces \cite{arXiv:1112.0573}, and logarithmic violation of the area law \cite{o3,arXiv:1112.2702}. The assessment of the theoretical ideas can be performed on heavy quarkonium systems which are suitable and useful environments.
Effective field theory techniques are needed to study such QCD systems, known as non-relativistic QCD (NRQCD). Because the velocity of the quark is $ v\ll 1$ the dynamics of the heavy quarks can be described by an effective non-relativistic field theory. In fact, NRQCD explains the effective theory of heavy quarkonia \cite{cas,sot,nor}.
In the NRQCD formalism, the heavy quarkonium spectroscopy and inclusive decays can be systematically described by starting from QCD \cite{cas,sot,nor}.  

Melting of quarkonia, such as $J/\psi$ and excited states in the medium, is one of the main experimental indications of QGP formation \cite{kar}. Because of this, the heavy quark potential study is significant for investigating QGP. Maldacena was the first person who calculated the heavy quark potential by using his AdS/CFT correspondence correctly \cite{Maldacena}. He concluded that the energy behaves as purely Coulombian in $ \mathcal{N}=4 $ SYM at zero temperature which is the property of the conformal gauge theories. Since then, the AdS/CFT correspondence has been used to calculate the heavy quark potential in different backgrounds many times. The heavy potential has been calculated in a background with a blackening function \cite{Rey1,Rey2}. Also, the potential has been obtained and investigated for, a curved spacetime \cite{32}, a strongly coupled non-conformal plasma with anisotropy \cite{35,Giataganas:2012zy,Giataganas:2013lga,Chakraborty:2012dt}, the Lifshitz backgrounds \cite{Danielsson:2009gi}, the Lifshitz backgrounds with hyperscaling violation \cite{37,38}, and some AdS/QCD models \cite{39,40}. 

One of the main mechanisms in suppression is color screening \cite{mat}, but another probably more important mechanism is the imaginary potential \cite{im1,im2}. At first, Noronha and Dumitru calculated the imaginary potential of a quarkonium by using the AdS/CFT correspondence for $\mathcal{N}=4$ SYM theory \cite{jon}. Thermal fluctuations arisen from interactions between the heavy quarks and the medium make the imaginary part of the potential. This kind of calculation of the imaginary potential has been used many times. Among these calculations, the following studies can be mentioned, a static \cite{sta1,sta2} and a moving quarkonium \cite{ali}, and finite 't Hooft coupling corrections \cite{fad}. We can also refer to \cite{fadi,25,ins,24,28,69,70}. In \cite{oth1}, other approaches have been introduced to study the imaginary part of the potential.

The pure QCD dose not have the conformal invariance, so it is an acceptable approach that the conformal invariance of the AdS$ _{5} $ spacetime is broken. To achieve this goal, one method that can be applied is hyperscaling violation metrics. Then, NRQCD can be investigated by Lifshitz and hyperscaling violation metrics, because these metrics are non-relativistic for $ z\neq 1 $ and $ \theta\neq 0 $. Regarding these considerations, it is interesting to study the real and imaginary parts of the potential in a hyperscaling violated Yang-Mills-dilaton metric background. In our previous papers \cite{38,69}, we calculated the real and imaginary parts of the potential in the hyperscaling violation Einstein-Maxwell-dilaton black brane charged under one U(1) gauge field, respectively.
In this paper, we develop and extend them to the hyperscaling violated Yang-Mills-dilaton black holes with non-Abelian gauge field.
Our purpose is to investigate the effects of the parameters of the Yang-Mills charge $ e_3 $, the dynamical exponent $ z $, and the hyperscaling violation parameter $ \theta $ on these potentials. We observe that, changing the values of these parameters affect on both the real and imaginary potentials.
Comparing the real and imaginary parts of the potential is interesting in studying the melting of the quarkonia, which we conduct in this paper for both potentials.

This paper is organized as follows. In section \ref{sec2}, we briefly review the hyperscaling violated Yang-Mills-dilaton black holes. Then in section \ref{sec3}, we investigate the effects of the dynamical exponent $ z $, the hyperscaling violation parameter $ \theta $ and the Yang-Mills charge $e_3$ on the real part of the potential. In section \ref{sec4}, the effects of these parameters on the imaginary part of the potential are studied. Finally, we summarize our results in the last section.

\section{Yang-Mills-dilaton black holes}\label{sec2}
Now, we review the hyperscaling violated Yang-Mills-dilaton black holes \cite{Mirza}. We start with the following (n+1)-dimensional action,
\begin{eqnarray}
&&S=\dfrac{1}{16\pi G}\int d^{n+1}x\sqrt{-g} \left (R-V(\Phi)-\dfrac{4}{n-1}\partial_{\mu} \Phi \partial^{\mu} \Phi-\dfrac{1}{4}\sum_{i=1}^{3}e^{-4\xi_{i}\Phi/(n-1)}F_{i}^{2}\right ),~~~~~~~\\ &&V(\Phi)=2\Lambda e^{\lambda\Phi},\nonumber
\end{eqnarray}
where $ \Phi $ and $ V(\Phi) $ are the dilaton field and its related potential, respectively, and $ \Lambda $ is the cosmological constant.  The coupling strength of $ \Phi $ and $ F^{2}_{i} $ are estimated by the constants $ \xi_{i} $.

This action admits the hyperscaling violated Yang-Mills-dilaton black hole solutions in the Einstein frame which is given by \cite{Mirza},
\begin{eqnarray}
&&ds_{_E}^{2}=r^{2\alpha}\left( -\dfrac{r^{2z}f(r)}{R^{2z}}dt^{2}+\dfrac{R^{2}}{r^{2}f(r)}dr^{2}+r^{2}d{\Omega}_{k}^{2}\right), ~~~~\alpha:=-\theta /(n-1), \label{ads1}\\
&&f(r)=1+\dfrac{R^{2}(n-1)(n-2)}{[(n-1)z-2\theta] [z+n-3-\theta]r^{2}}-\dfrac{m}{r^{z+n-\theta-1}}\\
&&+\left\{
\begin{array}{rl}
&\dfrac{(n-1)^{2}(n-2)~R^{2}~e_{3}^{2}~b^{2(z-1)-\frac{2\theta}{n-1}}}{(\theta-n+1)[(n-z-3)(n-1)-(n-5)\theta]~r^{2(z+1)-\frac{4\theta}{n-1}}}, ~~~~z\neq n-3-\frac{n-5}{n-1}\theta\nonumber\\
&\dfrac{(n-1)(n-2)R^{2}e_{3}^{2}b^{2(z-1)-\frac{2\theta}{n-1}}}{(\theta-n+1)~r^{2(z+1)-\frac{4\theta}{n-1}}}~ln(r),~~~~~~~~~~~~~~~~~~~~~~~~~~~z=n-3-\frac{n-5}{n-1}\theta 
\end{array} \right. \nonumber\\
&&\Phi(r)=\dfrac{1}{2}\sqrt{(n-\theta-1)[(z-1)(n-1)-\theta]}~ ln\left (\dfrac{r}{b}\right ),
\end{eqnarray}
where $ z $ shows the dynamical exponent, $ \theta $ denotes the hyperscaling violation parameter should be negative ($ \theta<0 $), $ R $ is the AdS radius,  $ m $ and $ b $ are the integration constants that $ m $ relates to the mass of the black hole. If one wants to arrive to the asymptotic behavior of the metric ($ f(r)=1 $ as $ r \rightarrow \infty $),  two conditions should be satisfied:  $ z+(n-1)(\alpha+1)>0 $ and $ 2z+4\alpha+2>0 $.
Also, to have a physically dual field theory,  the null energy condition  (NEC) must be checked, $ T_{\mu\nu}N^{\mu}N^{\nu}\geq 0 $. This condition leads to  $ (\alpha+1)(\alpha+z-1) > 0 $ that implies $ \alpha \ne -1 $ or $ \alpha \ne 1-z $.  

Solving equation $ f(r_{+})=0 $ leads to calculate the event horizon radius $ r_{+} $ and so one have,
\begin{eqnarray}
&&m(r_{+})=r^{z+n-\theta-1}_{+}+\dfrac{R^{2}(n-1)(n-2)}{[(n-1)z-2\theta]~(z+n-3-\theta)}~r^{z+n-\theta-3}_{+}\\
&&+\left\{
\begin{array}{rl}
&\dfrac{(n-1)^{2}(n-2)~R^{2}~e_{3}^{2}~b^{2(z-1)-\frac{2\theta}{n-1}}}{(\theta-n+1)[(n-z-3)(n-1)-(n-5)\theta]~r_{+}^{z+\theta-n+3-\frac{4\theta}{n-1}}},~~~~z\neq n-3-\frac{n-5}{n-1}\theta, \nonumber\\
&\dfrac{(n-1)(n-2)~R^{2}~e_{3}^{2}~b^{2(z-1)-\frac{2\theta}{n-1}}}{(\theta-n+1)~r_{+}^{z+\theta-n+3-\frac{4\theta}{n-1}}}~ln(r_{+}),~~~~~~~~~~~~~~~~~~~~~~~~~~~~~z=n-3-\frac{n-5}{n-1}\theta. 
\end{array} \right. 
\end{eqnarray}
Relating on what the values of the parameters of $ m, n, z, e_{3} $ and $ \theta $ are, the solution can be led to a black hole with two horizons, an extreme black hole, or a naked singularity.

The Hawking temperature is,
\begin{eqnarray}
T_{+}&=&\dfrac{r^{z+1}f'(r)}{4\pi~ R^{z+1}}|_{r=r_{+}}\\
&=&\dfrac{(z+n-\theta-1)r_{+}^{z}}{4\pi R^{z+1}}+\dfrac{(n-1)(n-2)r_{+}^{z-2}}{4\pi[(n-1)z-2\theta] ~R^{z-1}}+\dfrac{(n-1)(n-2) ~e_{3}^{2}~ b^{2(z-1)-\frac{2\theta}{n-1}}}{4\pi(\theta -n+1)R^{z-1}~r_{+}^{z+2-\frac{4\theta}{n-1}}}.~~~\nonumber
\label{TT} 
\end{eqnarray}
Also, the Yang-Mills charge of this black hole can be calculated from the Gauss law \cite{Mirza},
\begin{equation}
Q=\dfrac{1}{4\pi\sqrt{(n-1)(n-2)}}\int d^{n-1}x\sqrt{Tr\left (F_{\mu\nu}^{(a)}F_{\mu\nu}^{(a)}\right )}=\dfrac{\omega_{n-1}}{4\pi R^{1-z}}e_{3}.
\end{equation}

It is important to notice that geometries with nontrivial hyperscaling violation parameter can be viewed as an \textit{effective} holographic description of certain strongly coupled QFTs that are valid at energies below a specific scale in the theory. Therefore, the metric background (\ref{ads}) is not assumed to be valid for the whole way to the boundary. However, it accurately describes the dual field theory living on a background spacetime with a fixed surface value of $ r_{F} $. As a result, the QFT is effectively placed on the boundary of a UV cutoff.

\section{Real potential}\label{sec3}
In this section, we are going to investigate the real part of the potential in the hyperscaling violated Yang-Mills-dilaton black hole \cite{Mirza}. For this purpose, we consider the Wilson loop which is a significant and main observable in non-Abelian gauge theories   
\begin{equation}\label{key}
W(\mathcal{C})=\frac{1}{N_c}tr~\!\!P~\!exp\left[ig \oint_\mathcal{C} {\hat{A_{\mu}} dx^{\mu} }  \right], 	
\end{equation}
where the trace is taken over the fundamental representation of $ SU(N_c) $, $ P $ represents path-ordering, $ g $ denotes the coupling, $ \mathcal{C} $ is a closed loop in a 4-dimensional spacetime and $ \hat{A_{\mu}} $ shows the gauge potential. When  $ \mathcal{T}\rightarrow \infty $, the closed loop $ \mathcal{C} $ becomes the rectangular loop. In this limit, the expectation value of this rectangular Wilson loop yields the heavy quark potential
\begin{equation}\label{key}
	\left\langle W(\mathcal{C})\right\rangle \sim e^{-i\mathcal{T}V_{q\bar{q}}}. 	
\end{equation}
On the other hand, the expectation value of the Wilson loop from the holography viewpoint is obtained by,
\begin{equation}\label{key}
	\left\langle W(\mathcal{C})\right\rangle \sim e^{-iS_{\mathcal{C}}},
\end{equation}
where $ S_{\mathcal{C}} $ denotes the regularized action. Therefore, the heavy quark potential is given by,
\begin{equation}\label{key}
	V_{q\bar{q}}=\dfrac{S_{\mathcal{C}}}{\mathcal{T}}. 	
\end{equation}
To proceed, we calculate the heavy quark potential in (\ref{ads1}) background.     
For this purpose, we consider a rectangular Wilson loop which length of its short side, $ L $, puts along spatial extension and its  long side, $ \mathcal{T} $, puts along a time direction. A quark-antiquark pair is situated along the spatial direction so that the quarks are placed at $ x_{1}=\pm L/2 $.\\
In the string frame the background (\ref{ads1}) is given by
\begin{eqnarray}
	&&ds_{_S}^{2}=b^{-\beta}r^{2\alpha+\beta}\left( -\dfrac{r^{2z}f(r)}{R^{2z}}dt^{2}+\dfrac{R^{2}}{r^{2}f(r)}dr^{2}+r^{2}d{\Omega}_{k}^{2}\right), \label{ads}~~~~
\end{eqnarray}
where $\beta:=2\sqrt{(n - \theta - 1)[(z - 1)(n - 1) - \theta]}/3$.
The string worldsheet coordinates are parameterized as,
\begin{equation}
t=\tau,~~~~ x_{1}=\sigma,~~~~ x_{2}=0,~~~~x_{3}=0,~~~~r=r(\sigma).
\label{par} 
\end{equation}
One can start with the Nambu-Goto action,
\begin{equation}
S=-\dfrac{1}{2\pi\alpha^{\prime}}\int d\tau d\sigma \mathcal{L} =-\dfrac{1}{2\pi\alpha^{\prime}}\int d\tau d\sigma \sqrt{-g},\label{NG}
\end{equation}
where $ g $, the determinant of the induced metric of the string worldsheet, is given by,
\begin{equation}
g_{\rho\delta}=G_{\mu\nu}\partial_{\rho}X^{\mu} \partial_{\delta}X^{\nu}.
\end{equation} 
In the above equation, $ G_{\mu\nu} $ indicates the metric and $ X^{\mu} $ represents the target space coordinates.
Substituting (\ref{par}) into (\ref{ads}), the induced metric can be obtained as following,
\begin{eqnarray}
g_{00}&=&-\dfrac{r^{2\alpha+\beta +2z}}{R^{2z}} f(r),\nonumber\\
g_{11}&=&r^{2\alpha+\beta +2}\left( 1 +\dfrac{R^{2}r'^{2}}{r^{4}f(r)} \right), 
\end{eqnarray}
where the prime denotes the derivative with
respect to $ \sigma $. 
Then, the corresponding Lagrangian density can be calculated,
\begin{equation}
\mathcal{L}=\sqrt{M(r)+N(r)r'^{2}},\label{lag} 
\end{equation}
where,
\begin{eqnarray}
M(r)&=& \dfrac{r^{4\alpha+2\beta+2z+2}}{R^{2z}}f(r)=\dfrac{r^{\gamma}}{R^{2z}}f(r),~~~~~\gamma :=4\alpha+2\beta+2z+2, \label{a(r)}\\
N(r)&=& \dfrac{r^{4\alpha+2\beta+2z-2}}{R^{2z-2}}=\dfrac{r^{\gamma-4}}{R^{2z-2}}.\label{b(r)}
\end{eqnarray} 
The Hamiltonian $ H $ is a constant of the motion, because the Lagrangian (\ref{lag}) does not explicitly depend to $ \sigma $, then,
\begin{equation}
H=\mathcal{L}-\dfrac{\partial \mathcal{L}}{\partial r^{\prime }}r^{\prime}=constant.
\end{equation}
Due to the boundary condition, the string configuration is U-shaped and has a minimum at $ \sigma=0 $, i.e. $ r(0)=r_{c} $, so that $ r^{\prime}_{c}=0 $. Hence, we get,
\begin{equation}
r'=\sqrt{\dfrac{M^{2}(r)-M(r)M(r_{c})}{N(r)M(r_{c})}},\label{r'}
\end{equation}
where,
\begin{eqnarray}
&&M(r_{c})=\dfrac{r_{c}^{\gamma}}{R^{2z}}f(r_{c}) , \label{ac}\\
&&f(r_{c})=1+\dfrac{R^{2}(n-1)(n-2)}{[(n-1)z-2\theta] [z+n-3-\theta]~r_{c}^{2}}-\dfrac{m}{r_{c}^{z+n-\theta-1}}\\
&&+\left\{
\begin{array}{rl}
&\dfrac{(n-1)^{2}(n-2)~R^{2}~e_{3}^{2}~b^{2(z-1)-\frac{2\theta}{n-1}}}{(\theta-n+1)[(n-z-3)(n-1)-(n-5)\theta]~r_{c}^{2(z+1)-\frac{4\theta}{n-1}}}, ~~~~z\neq n-3-\frac{n-5}{n-1}\theta\nonumber\\
&\dfrac{(n-1)(n-2)~R^{2}~e_{3}^{2}~b^{2(z-1)-\frac{2\theta}{n-1}}}{(\theta-n+1)r^{2(z+1)-\frac{4\theta}{n-1}}}~ln(r_{c}).~~~~~~~~~~~~~~~~~~~~~~~~~~~z=n-3-\frac{n-5}{n-1}\theta 
\end{array} \right.
\end{eqnarray}
We can obtain the inter-distance $ L $ of the quark and antiquark by integrating from (\ref{r'}) as,
\begin{equation}
L=2\int dr \sqrt{\dfrac{N(r)M(r_{c})}{M^{2}(r)-M(r)M(r_{c})}}.\label{LL}
\end{equation}
Substituting (\ref{r'}) into (\ref{NG}) gives the action of the quark-antiquark pair,
\begin{equation}
S=\dfrac{\mathcal{T} }{\pi\alpha^{\prime}}\int  dr \sqrt{\dfrac{M(r)N(r)}{M(r)-M(r_{c}) }}. \label{11}
\end{equation}
The free quark and antiquark contributions cause to diverge this action. Therefore, one should subtract the self-energy of two quarks from (\ref{11}) to remove this divergency. The self-energy is,
\begin{equation}
S_{0}=\dfrac{\mathcal{T} }{\pi\alpha^{\prime}}\int dr \sqrt{N(r_{0})},
\end{equation}
where $ N(r_{0})=N(r\rightarrow \infty) $. Then, the real part of the heavy quark potential is given by,
\begin{equation}
ReV_{Q\overline{Q}}=\dfrac{1}{\pi\alpha'}\int_{r_{c}}^{r_{F}} dr \left( \sqrt{\dfrac{M(r)N(r)}{M(r)-M(r_{c}) }}-\sqrt{N(r_{0})}\right)-\dfrac{1}{\pi\alpha'}\int_{r_{+}}^{r_{c}}  dr \sqrt{N(r_{0})}.
\end{equation}

\begin{figure}
	\begin{center}
		\includegraphics[ width=73 mm]{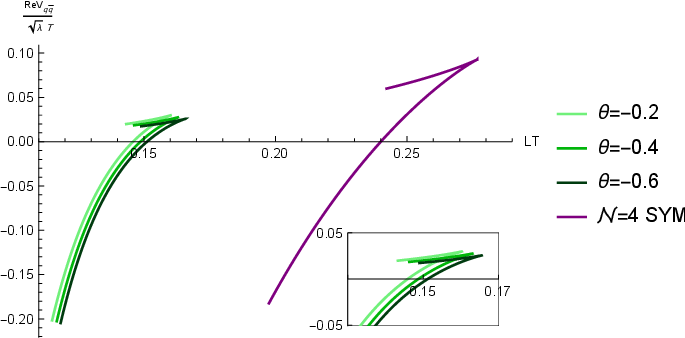}
	\end{center}
	\caption{ $ V_{q\bar{q}} $ against $ LT $ in the different $ \theta $ at $ z=2 $ and $ e_3=1 $.}
	\label{Vtheta}
\end{figure}
\begin{figure}
	\begin{center}
		\includegraphics[width=73 mm]{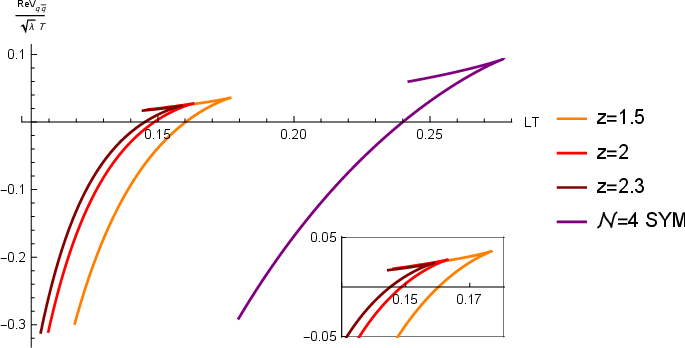}
	\end{center}
	\caption{ $ V_{q\bar{q}} $ against $ LT $ in the different $ z $ at $\theta=-0.4 $ and $ e_3=1 $.}
	\label{Vz}
\end{figure}
\begin{figure}
	\begin{center}
		 \includegraphics[width=73 mm]{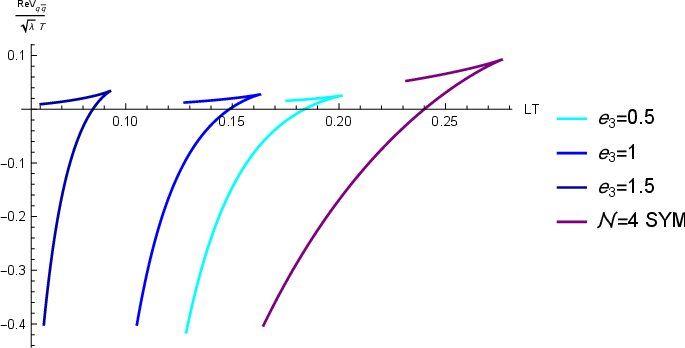}
	\end{center}
	\caption{ $ V_{q\bar{q}} $ against $ LT $ in the different $ e_3 $ at $ \theta=-0.4 $ and $ z=2 $.}
	\label{Ve}
\end{figure}
To investigate the effects of the hyperscaling violation parameter ($ \theta $), the dynamical exponent ($ z $), and the Yang-Mills charge ($ e_3 $) on the real part of potential, we have drawn $ \frac{V}{\sqrt{\lambda}T} $\footnote{The 't Hooft coupling of the gauge theory is given by $ \sqrt{\lambda}=R^{2}/\alpha'. $} as a function of $ LT $ in Figs \ref{Vtheta},\ref{Vz} and \ref{Ve} \footnote{We consider $ R=1 $, $ r_{+}=1 $, $ n=4 $ and $ b=1 $ in all plots.}. Indeed, the real potential has a complex dependency on $ r $ at the different values of $ z $ and $ \theta $, so it can not be solved analytically and numerical methods must be used.

As seen clearly from Fig \ref{Vtheta}, increasing $ \theta $ increases the real potential and decreases the dissociation length at the fixed $ z $ and $ e_3 $. Note that, the real potential in the $ \mathcal{N}=4 $ SYM is less than those in the hyperscaling violated Yang-Mills-dilaton background, and its dissociation length is longer. Therefore, the quarkonium dissociates harder in the $ \mathcal{N}=4 $ SYM than in the hyperscaling violated Yang-Mills-dilaton background. 

In Fig \ref{Vz}, we have studied the effect of $ z $ on $ V $  at the fixed $ \theta $ and $ e_3 $. It shows that the real potential tends to be increased with increasing $ z $, thus the dissociation length decreases. The quarkonium dissociates easier in the hyperscaling violated Yang-Mills-dilaton background than in the $ \mathcal{N}=4 $ SYM. Also, the static potential was investigated in strongly coupled anisotropic $ \mathcal{N}=4 $ super Yang-Mills plasma in \cite{Giataganas:2012zy,Giataganas:2013lga,35,Chakraborty:2012dt,Rougemont:2014efa}. As they showed the anisotropic parameter decreases the absolute value of the static potential and the dissociation length. The manner of acting of the anisotropic parameter $a$ is similar to the dynamical exponent $z$ here.

Finally, we have drawn $ V $ for different values of $ e_3 $ at the fixed $ \theta $ and $ z $ in Fig \ref{Ve}. As seen clearly, in the presence of the Yang-Mills charge the real potential increases, hence the dissociation length decreases.
Note that, the quarkonium dissociation is easier in the hyperscaling violated Yang-Mills dilaton background than in $ \mathcal{N}=4 $ SYM.

As a result, $ \theta $, $ z $, and $ e_3 $ tend to make the  melting of the heavy quarkonium easier.

The behavior of the heavy quark potential was investigated in two different hyperscaling violation metric background in \cite{38,37}. They concluded that the real potential increases by increasing $z$ and decreasing $\theta$. However, we obtained that increasing $z$ and $\theta$ leads to increasing the real potential. As can be concluded, the eﬀect of the dynamical exponent $z$ is in agreement, but the hyperscaling violation parameter $\theta$ has an opposed effect. This disagreement arises from discrepancies in the metric backgrounds.

\section{Imaginary potential}\label{sec4}

Here, we investigate the imaginary part of the heavy quark potential in the hyperscaling violated Yang-Mills-dilaton black hole \cite{Mirza}, by applying the thermal worldsheet fluctuations method introduced in Refs. \cite{sta1,jon}. It reads, 
\begin{equation}
ImV_{Q\overline{Q}}=-\dfrac{1}{2\sqrt{2}\alpha'}\sqrt{N(r_{c})}\left(\dfrac{M'(r_{c})}{2M''(r_{c})}-\dfrac{M(r_{c})}{M'(r_{c})}\right),\label{imm} 
\end{equation}
where,
\begin{eqnarray}
M(r_{c})&=&\dfrac{r_{c}^{\gamma}}{R^{2z}}f(r_{c}) , \label{ac}\\	
N(r_{c})&=&\dfrac{r_{c}^{\gamma-4}}{R^{2z-2}},\\
M'(r_{c})&=&\dfrac{r_{c}^{\gamma}}{R^{2z}}  \left(f'_{c}+\dfrac{\gamma}{r_{c}}f_{c} \right),\\
M''(r_{c})&=&\dfrac{r_{c}^{\gamma}}{R^{2z}} \left(f''_{c}+\dfrac{2\gamma}{r_{c}}f'_{c}+\dfrac{\gamma(\gamma-1)}{r_{c}^{2}}f_{c} \right),~~~~~~~~~~
\end{eqnarray}
and,
\begin{eqnarray}
f'(r_{c})&=& -\dfrac{2R^{2}(n-1)(n-2)}{[(n-1)z-2\theta] [z+n-3-\theta]r^{3}}+\dfrac{(z+n-\theta-1)m}{r^{z+n-\theta}}\\
&-&\left\{
\begin{array}{rl}
&\dfrac{[2(z+1)-\frac{4\theta}{n-1} ](n-1)^{2}(n-2)~R^{2}~e_{3}^{2}~b^{2(z-1)-\frac{2\theta}{n-1}}}{(\theta-n+1)[(n-z-3)(n-1)-(n-5)\theta]~r^{2z+3-\frac{4\theta}{n-1}}}, \nonumber\\
&\dfrac{(n-1)(n-2)~R^{2}~e_{3}^{2}~b^{2(z-1)-\frac{2\theta}{n-1}}}{(\theta-n+1)~r^{2z+3-\frac{4\theta}{n-1}}}~[(2(z+1)-\dfrac{4\theta}{n-1} )ln(r)-1], 
\end{array} \right. \label{ffc}
\end{eqnarray}
\begin{eqnarray}
f''(r_{c})&=&\dfrac{6R^{2}(n-1)(n-2)}{[(n-1)z-2\theta] [z+n-3-\theta]r^{4}}-\dfrac{(z+n-\theta-1)(z+n-\theta)m}{r^{z+n-\theta+1}}\\
&+&\!\!\left\{
\begin{array}{rl}
&\!\!\!\dfrac{ [2(z+1)-\frac{4\theta}{n-1} ] [2z+3-\frac{4\theta}{n-1} ](n-1)^{2}(n-2)R^{2}~e_{3}^{2}~b^{2(z-1)-\frac{2\theta}{n-1}}}{(\theta-n+1)[(n-z-3)(n-1)-(n-5)\theta]~r^{2z+4-\frac{4\theta}{n-1}}}, \nonumber\\
&\!\!\!\dfrac{ [2z+2-\frac{4\theta}{n-1}] (n-1)(n-2)R^{2}~e_{3}^{2}~b^{2(z-1)-\frac{2\theta}{n-1}}}{(\theta-n+1)~r^{2z+4-\frac{4\theta}{n-1}}}~[ (2(z+1)-\dfrac{4\theta}{n-1}+1)ln(r)-2],
\end{array} \right. \label{fff}
\end{eqnarray}
where the derivatives are with respect to $ r $.

Next, based on what has been mentioned in \cite{jon,fi}, three restrictions on the formula (\ref{imm}) should be imposed. The first restriction is that the term $ N(r_{c}) $ must be positive. Therefore, we have,
\begin{equation}
\dfrac{r_{c}^{\gamma -4}}{R^{2z-2}}> 0,
\end{equation}
as seen clearly it is always positive.

The second restriction is that the imaginary part of potential should be negative,
\begin{equation}
\dfrac{M'(r_{c})}{2M''(r_{c})}-\dfrac{M(r_{c})}{M'(r_{c})} >0.
\end{equation}

\begin{figure}
\begin{center}$
\begin{array}{cc}
\includegraphics[width=73 mm]{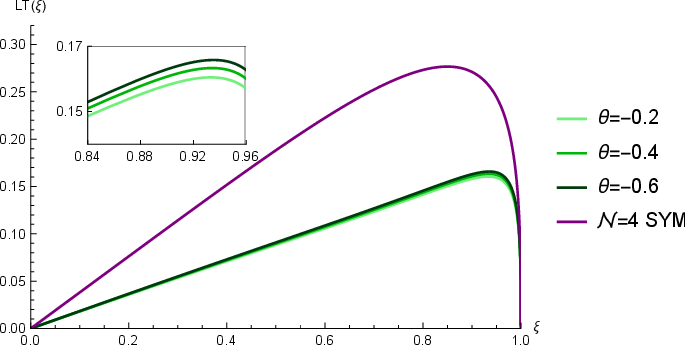}
\hspace{0.2cm}
\includegraphics[width=73 mm]{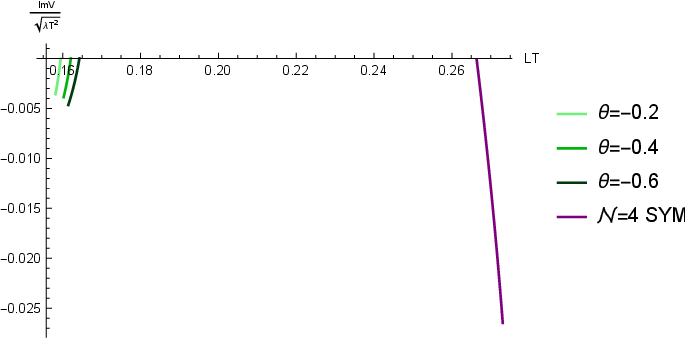}
\end{array}$
\end{center}
\caption{Left: $ LT $ against $ \xi $ in the different $ \theta $ at $ z=2 $ and $ e_3=1 $. Right: $ ImV_{Q\overline{Q}}/\sqrt{\lambda}T $ against $ LT $ in the different $ \theta $ at $ z=2 $ and $ e_3=1 $.}
\label{ww}
\end{figure}

\begin{figure}
\begin{center}$
\begin{array}{cc}
\includegraphics[width=73 mm]{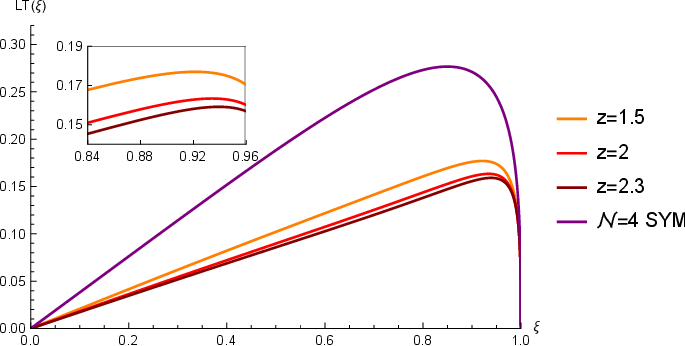}
\hspace{0.2cm}
\includegraphics[width=73 mm]{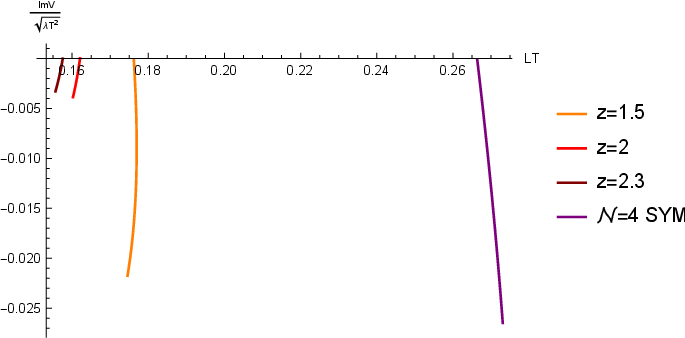}
\end{array}$
\end{center}
\caption{Left: $ LT $ against $ \xi $ in the different $ z $ at $ \theta=-0.4 $ and $ e_3=1 $. Right: $ ImV_{Q\overline{Q}}/\sqrt{\lambda}T $ against $ LT $ in the different $ z $ at $ \theta=-0.4 $ and $ e_3=1 $.}
\label{w}
\end{figure}

\begin{figure}
\begin{center}$
\begin{array}{cc}
\includegraphics[width=73 mm]{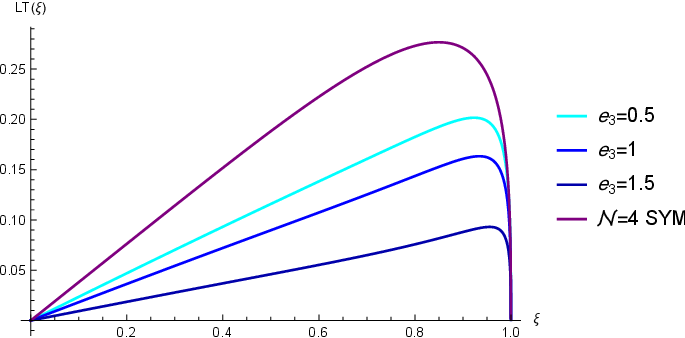}
\hspace{0.2cm}
\includegraphics[width=73 mm]{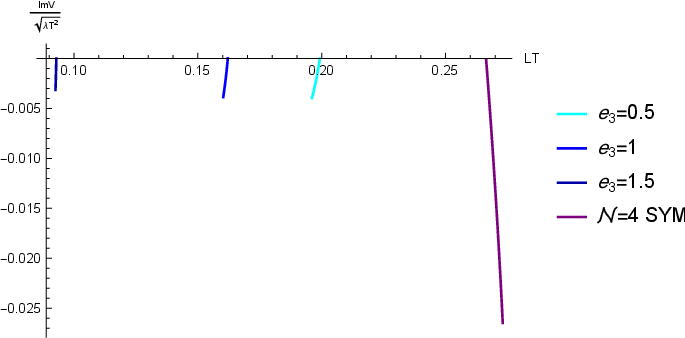}
\end{array}$
\end{center}
\caption{Left: $ LT $ against $ \xi $ in the different $ e_3 $ at $ \theta=-0.4 $ and $ z=2 $. Right: $ ImV_{Q\overline{Q}}/\sqrt{\lambda}T $ against $ LT $ in the different $e_3 $ at $ \theta=-0.4 $ and $ z=2 $.}
\label{www}
\end{figure}

The third restriction is related to the determination of the maximum value of the $ LT(\xi) $, $ LT_{max} $. To address this, we have drawn numerically $ LT $ versus $ \xi(\equiv r_{+}/r_{c}) $ for some choices $ z $, $ \theta $ and $ e_{3} $ in figures \ref{ww}, \ref{w} and \ref{www}. As seen clearly, by increasing $ \xi $ up to the critical value $ \xi_{max} $, $ LT $ rises insofar as it reaches $ LT_{max} $ and thereafter descends by increasing $ \xi $. Note that, in the region of $ \xi >\xi_{max} $, highly curved configurations for the string worldsheet that are not solutions of the Nambu-Goto action, are dominated \cite{bac}. Therefore, we should consider just the region of $ \xi < \xi_{max} $.

Here, we purpose to study the effects of the hyperscaling violation parameter $ \theta $, the dynamical exponent $ z $ and the charge $ e_{3} $ on the inter-distance and the imaginary potential. Due to the  complex dependency of the inter-distance and the imaginary potential on $ r $ at the different values of $ z $ and $ \theta $, they  can not be solved analytically and numerical methods must be used.

We draw numerically $ LT $ versus $ \xi $ and $ ImV_{Q\overline{Q}}/(\sqrt{\lambda}T) $ against $ LT $ for the different values of $ \theta $, in figure \ref{ww}. As seen in the left panel, $ LT_{max} $ decreases by increasing $ \theta $ at the fixed $ z $ and $ e_{3} $. 
From \cite{sta1} we know that in the case of $ \mathcal{N}=4 $ SYM, $ LT_{max}=LT(\xi_{max})\sim 0.28 $. We may note that, $ LT_{max} $ in $\mathcal{N}=4 $ SYM is bigger than the corresponding values of $ LT_{max} $ at the different  values of $ \theta $. Also, in the right panel, the imaginary potential commences at $ L_{min} $ which is the solution of the equation $ ImV_{Q\overline{Q}}=0 $,  and finishes at $L_{max}$. By increasing $ \theta $ at the fixed $ z $ and $ e_{3} $, the absolute value of the imaginary potential decrease. Actually, the hyperscaling violation parameter $ \theta $ gives rise to $ ImV_{Q\overline{Q}}/(\sqrt{\lambda}T) $ at the smaller distances. As mentioned in \cite{fi},the suppression is stronger if $ ImV_{Q\overline{Q}}/(\sqrt{\lambda}T) $ starts at the smaller $ LT $. So, $ \theta $ tends to decrease the thermal width and makes the suppression stronger. Also, the quarkonium dissociates harder in the $ \mathcal{N}=4 $ SYM than in the hyperscaling violated Yang-Mills-dilaton black hole.

As seen from figure \ref{w}, the $ LT $ and $ ImV_{Q\overline{Q}}/(\sqrt{\lambda}T) $ have been plotted numerically against $ \xi $ and $ LT $, respectively, for the different values of $ z $ at the fixed $ \theta $ and $ e_{3} $. From the left panel, one can see that the values of $ LT_{max} $ decreases as $ z $ increases. Also, these are less than the corresponding $ LT_{max} $ in $\mathcal{N}=4 $ SYM. In the right panel, by increasing $ z $,  the absolute value of the imaginary potential decreases and the onset of $ ImV_{Q\overline{Q}}$ happens at the smaller $LT$. The dynamical exponent $ z $  tends to decrease $ LT_{max} $ and the onset of the imaginary potential, therefore strengthens the suppression. The quarkonium dissociation is easier in the hyperscaling violated Yang-Mills-dilaton black hole than in $ \mathcal{N}=4 $ SYM.
 
Also, the imaginary potential has been studied in strongly coupled anisotropic $ \mathcal{N}=4 $ SYM plasma in \cite{BitaghsirFadafan:2013vrf,Giataganas:2013lga}. As they showed the anisotropic parameter decreases the imaginary potential. The behavior of anisotropic parameter $a$ is similar to the dynamical exponent $z$ here. 
 
We have plotted $ LT $ and $ ImV_{Q\overline{Q}}/(\sqrt{\lambda}T) $ against $ \xi $ and $ LT $ respectively, for the different values of $ e_{3} $, in figure \ref{www}. As seen clearly, $ LT_{max} $ grows as $ e_{3} $ lowers at the fixed $ z $ and $ \theta $. 
These corresponding $ LT_{max} $s are less than that of $\mathcal{N}=4 $ SYM.
Moreover, as $ e_{3} $ increases, the absolute value of the imaginary potential and its onset decrease. So, the Yang-Mills charges decrease the thermal width and strengthen the suppression. Note that, the quarkonium dissociation is easier in the hyperscaling violated Yang-Mills dilaton background than in $ \mathcal{N}=4 $ SYM. 

The imaginary part of the potential was studied in the hyperscaling violating metric background at the Einstein-Maxwell-Dilaton theory in \cite{69}. They concluded that the dynamical exponent $z$ and the hyperscaling violation parameter $\theta$ behave differently. As $z$ increases the onset of the imaginary potential decreases and makes the suppression stronger, while  $\theta$ behaves opposite. Our result about $z$ is in agreement with \cite{69}, while the result of $\theta$ is in disagreement. This inconsistency appears from the difference in the metric backgrounds, as concluded in the real potential section too.

Also, by substituting $ z=1 $, $ \theta =0 $ and $ e_3=0 $ into (\ref{imm}), we recovered the result of \cite{jon}. 

\section{Conclusion}\label{con}
The dissociation of a heavy quarkonium in heavy ion collisions is an important experimental signature confirming the formation of QGP. Two approaches, the real potential and the imaginary potential, are employed to investigate the melting of the heavy quarkonium. We set out to study the effects of the dynamical exponent $ z $, the Yang-Mills charge $ e_3 $, and the hyperscaling violation parameter $ \theta $ on the real and imaginary parts of the potential in the hyperscaling violated Yang-Mills-dilaton metric background.

For the real part of the potential,  we conclude that increasing $ z $, $ \theta $, and $ e_3 $ lead to increasing the real potential and decreasing the dissociation length. Therefore, the dissociation of the quarkonium becomes easier. For $\theta$, $z$, and $e_3$, the real potential in $\mathcal{N}=4$ SYM becomes less than in the hyperscaling violated Yang-Mills dilaton background. So, the quarkonium in $\mathcal{N}=4$ SYM dissociates harder.  

For the imaginary part of the potential, we conclude that $ z $, $ \theta $, and $ e_3 $ tend to decrease the thermal width and the onset of the imaginary potential,  therefore make the suppression stronger. Thus, the dynamical exponent, the hyperscaling violation parameter, and the Yang-Mills charge make the quark-antiquark melting easier. Also, at $ z $, $ \theta $, and  $ e_3 $, the quarkonium dissociation gets easier in the hyperscaling violated Yang-Mills-dilaton black hole than in the $ \mathcal{N}=4 $ SYM.  Furthermore, the outcomes obtained from the real and imaginary potential approaches are compatible.

The anisotropy effect has been investigated on the real potential in \cite{Giataganas:2012zy,Giataganas:2013lga,Chakraborty:2012dt,Rougemont:2014efa,35} and the imaginary potential in \cite{BitaghsirFadafan:2013vrf,Giataganas:2013lga}. As can be seen from these references, the anisotropy tends to decrease the absolute value of the real potential and the onset of the imaginary potential.
These results are in accordance with our results of the dynamical exponent $z$, although their metric backgrounds are different from our metric background.  
Therefore, the dynamical exponent $z$ as an anisotropy parameter can make the quarkonium dissociation easier. 
  
We conclude that the Yang-Mills charge $e_3$ tends to increase and decrease the real potential and the imaginary potential, respectively. This can be expected due to the effect of the medium on the quarkonium. In other words, as the charge increases the screening effects of the medium can enhance. 

It is worth noting to mention that the metric background we consider here is a hyperscaling violated Yang-Mills-dilaton black hole. This is the extent of the dilaton Yang-Mills black hole solutions with a hyperscaling violation. The main difference of this black hole from the usual hyperscaling black holes (i.e. the different form of $f(r)$ blackening factor) is the appearance of non-Abelian fields (three fields) instead of Abelian fields. In the case $n>3$ the mentioned black hole is different from the hyperscaling violation Maxwell-dilaton theory. In fact, this black hole reduces to the Maxwell theory for $n=3$. This hyperscaling violated spacetime divides into two categories $z=n-3-\frac{n-5}{n-1}\theta$ and $z\neq n-3-\frac{n-5}{n-1}\theta$, while the Maxwell-Lifshitz theory contains only one category for each value of $z$. Also, this solution is appropriate only for the spherical hypersurface in the hyperscaling violated spacetime. For these reasons, our quantitative results are different from the other references.

\hfill \\ \\
\textbf{Acknowledgments}\\
We would like to thank B. Mirza for inspiring discussions. We are also grateful to M. Khanpour and N. Kiomarsipour for editing of this manuscript.

\end{document}